\documentstyle[aps,preprint,prl,epsfig]{revtex}
\renewcommand{\baselinestretch}{1.7}
\newcommand{\ba}{\begin{eqnarray}}
\newcommand{\ea}{\end{eqnarray}}
\begin{document}

\title{Generalized Partial Dynamical Symmetry in Nuclei}
\author{A. Leviatan$^{1}$ and P.~Van Isacker$^{2}$}
\address{$^{1}$Racah Institute of Physics, The Hebrew University,
Jerusalem 91904, Israel}

\address{$^{2}$Grand Acc\'el\'erateur National d'Ions Lourds,
B.P.~55027, F-14076 Caen Cedex 5, France}
\date{\today}
\maketitle

\begin{abstract}
We introduce the notion of a generalized partial dynamical symmetry
for which part of the eigenstates have part of the dynamical symmetry.
This general concept is illustrated with the example
of Hamiltonians with a partial dynamical $O(6)$ symmetry
in the framework of the interacting boson model.
The resulting spectrum and electromagnetic transitions are compared with
empirical data in $^{162}$Dy.\\
\end{abstract}


\newpage

The concept of dynamical symmetry has been widely used
in diverse areas of physics
with notable examples in nuclear, molecular,
and hadronic physics~[1-3].
In this approach one assumes that the Hamiltonian
can be written in terms of the Casimir operators
of a chain of nested algebras
\begin{equation}
G_1 \supset G_2 \supset \cdots \supset G_n ~,
\label{ds}
\end{equation}
in which case it has the following properties:
(i) {\it Solvability:}
all states are solvable
and analytic expressions
are available for energies and other observables;
(ii) {\it Quantum numbers:}
all states are classified by quantum numbers
which are the labels of irreducible representations (irreps)
of the algebras in the chain;
(iii) {\it Pre-determined structure:}
the structure of wave functions is completely dictated by symmetry
and is independent of the Hamiltonian's parameters.
The merits of a dynamical symmetry are self-evident.
However, in most applications to realistic systems,
the predictions of an exact dynamical symmetry are rarely fulfilled
and one is compelled to break it. This is usually done
by including in the Hamiltonian symmetry-breaking terms associated with
different sub-algebra chains of the parent spectrum generating algebra
($G_1$). In general, under such circumstances, solvability is lost,
there are no remaining non-trivial conserved quantum numbers and all
eigenstates are expected to be mixed.
A partial dynamical symmetry (PDS) corresponds to
a particular symmetry breaking for which some (but not all) of the above
mentioned virtues of a dynamical symmetry are retained.
Such intermediate symmetry structures were recently shown
to be relevant for nuclear~[4-10] and molecular~\cite{pinchen97}
spectroscopy, as well as to the study of mixed systems with coexisting
regularity and chaos~\cite{whelan93}.

Two types of PDS were encountered so far.
The first type corresponds to a situation for which
{\bf part} of the states preserve {\bf all} the dynamical symmetry.
In this case the properties of solvability, good quantum numbers,
and symmetry-dictated structure are fulfilled
exactly, but by only a subset of states. An example in the framework of the
interacting boson model (IBM-1)~\cite{ibm} is the chain
$U(6) \supset SU(3) \supset O(3)$, applicable to axially deformed nuclei,
where a non-$SU(3)$-scalar Hamiltonian has been constructed and shown to have
a subset of solvable states with good $SU(3)$ symmetry while other
states are mixed~\cite{lev96,levsin99}.

The second type of PDS corresponds to a situation for which {\bf all} the
states preserve {\bf part} of the dynamical symmetry.
In this second case there are no analytic solutions,
yet selected quantum numbers (of the conserved symmetries) are retained.
This occurs, for example, when the Hamiltonian contains interaction
terms from two different chains with
a common symmetry subalgebra, {\it e.g.} the $U(5)\supset O(5)$  and
$O(6)\supset O(5)$ chains in the IBM-1~\cite{levnov86}.
Alternatively, this type of PDS occurs when the Hamiltonian
preserves only some of the symmetries $G_i$ in the chain~(\ref{ds}) and
only their irreps are unmixed.
Such a scenario was recently
considered in~\cite{isa99} in relation to the chain
\begin{equation}
\begin{array}{ccccccc}
U(6) &\supset& O(6) &\supset& O(5) &\supset& O(3) \\
 \left [N\right ] && \langle0,\sigma,0\rangle && (\tau,0) && L
\end{array}.
\label{dsO6}
\end{equation}
An IBM-1 Hamiltonian was constructed which preserves the
$U(6)$, $O(6)$, and $O(3)$ symmetries (with quantum numbers $N,\sigma,L$)
but not the $O(5)$ symmetry (and hence leads to $\tau$ admixtures).
To obtain this type of PDS in the IBM-1, it is necessary to
include higher-order (three-body) terms in the Hamiltonian.

The purpose of the present work is
to show that it is possible to combine both types of PDS, namely,
to construct a Hamiltonian for which {\bf part} of the
states have {\bf part} of the dynamical symmetry.
We refer to such a  structure
as a generalized partial dynamical symmetry.
For the chain~(\ref{dsO6}) this can be achieved
with an IBM-1 Hamiltonian with only two-body interactions.
We analyze the resulting band structure and multi-phonon admixtures,
and compare the spectrum and E2 rates with empirical data in $^{162}$Dy.

The following type of IBM-1 Hamiltonian has been proposed~\cite{isa99}
as a representative of a PDS of the second kind
\begin{equation}
H_1 \;=\; \kappa_{0}P^{\dagger}_{0}P_{0}
+ \kappa_2 \Bigl (\Pi^{(2)}\times \Pi^{(2)}\Bigr )^{(2)}\cdot\Pi^{(2)} ~.
\label{h1}
\end{equation}
The $\kappa_0$ term is the $O(6)$ pairing term
defined in terms of monopole ($s$) and quadrupole ($d$) bosons,
$P^{\dagger}_{0}=  d^{\dagger}\cdot d^{\dagger} - (s^{\dagger})^2$.
It is diagonal in the dynamical-symmetry basis
$\vert [N],\sigma,\tau,L\rangle$ of Eq.~(\ref{dsO6}) with eigenvalues
$\kappa_0(N - \sigma)(N +\sigma +4)$.
The $\kappa_2$ term is constructed only from the $O(6)$ generator,
$\Pi^{(2)}=d^{\dagger}s+s^{\dagger}\tilde{d}$,
which is not a generator of $O(5)$.
Therefore, it cannot connect states in different $O(6)$ irreps
but can induce $O(5)$ mixing subject to $\Delta\tau=\pm 1,\pm 3$.
Consequently, all eigenstates of $H_1$
have good $O(6)$ quantum number $\sigma$
but do not possess $O(5)$ symmetry $\tau$.

To consider a generalized $O(6)$ PDS,
we introduce the following IBM-1 Hamiltonian:
\begin{equation}
H_2 \;=\; h_{0}P^{\dagger}_{0}P_{0} + h_{2}P^{\dagger}_{2}
\cdot\tilde P_{2} ~.
\label{h2}
\end{equation}
The $h_0$ term is identical to the $\kappa_0$ term of Eq.~(\ref{h1}),
and the $h_2$ term is defined in terms of the boson pair
$P^{\dagger}_{2,\mu} = \sqrt{2}\,s^{\dagger}d^{\dagger}_{\mu}
+ \sqrt{7}(d^{\dagger}d^{\dagger})^{(2)}_{\mu}$ with
$\tilde P_{2,\mu} = (-)^{\mu}P_{2,-\mu}$.
The multipole form of $H_2$ is
\ba
H_2 &=& h_{0}\left [-\hat C_{O(6)} + \hat N (\hat N+4)\right ]
+ h_{2}\,2\hat C_{O(5)} - h_{2}\hat C_{O(3)}
\nonumber\\
&& +\, h_2\,2\hat{n}_d (\hat N - 2)
+ h_{2}\sqrt{14}\,\Pi^{(2)}\cdot (d^{\dagger}\tilde d\,)^{(2)} ~,
\label{mult}
\ea
where $\hat N$ and $\hat{n}_d$ are the total and $d$-boson number
operators,
and $\hat C_{G}$ denotes
the quadratic Casimir operator of $G=O(6),\,O(5),\,O(3)$
with eigenvalues
$\sigma(\sigma+4),\,\tau(\tau+3),\,L(L+1)$, respectively.
The first three terms in Eq.~(\ref{mult})
are diagonal in the dynamical symmetry basis of Eq.~(\ref{dsO6}).
The $\hat{n}_d (\hat N - 2)$ term is a scalar under $O(5)$
but can connect states differing by $\Delta\sigma=0,\pm 2$.
The last term in Eq.~(\ref{mult}) induces both $O(6)$ and $O(5)$ mixing
subject to $\Delta\sigma=0,\pm 2$ and $\Delta\tau=\pm 1,\pm 3$.
Although $H_2$ is not an $O(6)$ scalar, it has an exactly
solvable ground band with good $O(6)$ symmetry. This
arises from the fact that the $O(6)$ intrinsic state for the
ground band
\ba
\vert c;\,N \rangle = (N!)^{-1/2}(b^{\dagger}_{c})^{N}\vert 0 \rangle
\;, \;\;\;
b^{\dagger}_c = (d^{\dagger}_{0} +s^{\dagger} )/\sqrt{2} ~,
\label{cond}
\ea
has $\sigma=N$ and is an exact zero energy eigenstate of $H_2$.
Since $H_2$ is rotational invariant, states of good angular momentum $L$
projected from $\vert c;\,N\rangle$ are also zero-energy eigenstates of
$H_2$ with good $O(6)$ symmetry, and form the ground band of
$H_2$. These projected states do not have good $O(5)$ symmetry and their
known wave functions contain a mixture of components with different $\tau$.
For example, the expansions of the ground state $L=0^{+}_{K=0_1}$ and
first excited state ($L=2^{+}_{K=0_1}$) of $H_2$ in the $O(6)$ basis
$\vert [N],\sigma,\tau,L\rangle$ have the form
\ba
\vert 0^{+}_{K=0_1}\rangle &=&
{\cal N}\sum_{n}a_n\vert\,[N],N,3n,0\, \rangle ~,
\nonumber\\
\vert 2^{+}_{K=0_1}\rangle &=&
{\cal N'}\sum_{n}\Bigl\{b_n\vert\,[N],N,3n+1,2\, \rangle
+ c_n\vert\,[N],N,3n+2,2\, \rangle \Bigr \} ~.
\label{02}
\ea
Here ${\cal N}$ and ${\cal N'}$ are normalization coefficients
and the amplitudes $a_n,\,b_n,\,c_n$ $(n=0,1,\ldots)$ are given by
$a_{n} = (-1)^{n}\sqrt{2n+1\over (N-3n)!(N+3n+3)!}$, 
$b_{n} = (-1)^{n}\sqrt{n+1\over (N-3n-1)!(N+3n+4)!}$ and 
$c_{n} = (-1)^{n+1}\sqrt{n+1\over (N-3n-2)!(N+3n+5)!}$.
It follows that $H_2$
has a subset of solvable states with good $O(6)$ symmetry ($\sigma=N$),
which is not preserved by other states.
All eigenstates of $H_2$ break the $O(5)$ symmetry but preserve the
$O(3)$ symmetry. These are precisely the required features of a generalized
PDS as defined above for the chain of Eq.~(\ref{dsO6}).

In Fig.~1 we show the experimental spectrum of $^{162}$Dy
and compare with the calculated spectra of $H_1$ and $H_2$.
The spectra display rotational bands of an axially-deformed nucleus,
in particular, a ground band $(K=0_1)$
and excited $K=2_1$ and $K=0_2$ bands.
An $L^2$ term is added to both Hamiltonians,
which contributes to the rotational splitting
but has no effect on wave functions.
The parameters are chosen to reproduce
the excitation energies
of the $2^+_{K=0_1}$, $2^+_{K=2_1}$, and $0^+_{K=0_2}$ levels.
The $O(6)$ decomposition of selected bands is shown in Fig.~2.
For $H_2$, the solvable $K=0_1$ ground band has $\sigma=N$
and exhibits an exact $L(L+1)$ splitting.
The $K=2_1$ band is almost pure with only $0.15\%$ admixture
of $\sigma=N-2$ into the dominant $\sigma=N$ component.
The $K=0_2$ band has components with $\sigma=N$ $(85.50\%)$,
$\sigma=N-2$ $(14.45\%)$, and $\sigma=N-4$ $(0.05\%)$.
These are the admixtures for the $K=2_1$ and $K=0_2$ bandheads;
they do not vary much throughout the bands
as long as the spin is not too high.
Higher bands exhibit stronger mixing,
{\it e.g.}, the $L=3^{+}$ member of the $K=2_3$ band shown 
in Fig.~2, has components 
with $\sigma=N$ $(50.36\%)$, $\sigma=N-2\,(49.25\%)$,
$\sigma=N-4\,(0.38\%)$ and $\sigma=N-6\,(0.01\%)$. 
The $O(6)$ mixing in excited bands of $H_2$
depends critically on the ratio $h_2/h_0$ in Eq.~(\ref{h2})
or equivalently on the ratio of the $K=0_2$ and $K=2_1$ bandhead energies.
In contrast, all bands of $H_1$ are pure with respect to $O(6)$.
Specifically, the $K=0_1,2_1,2_3$ bands shown in Fig.~2 have $\sigma=N$
and the $K=0_2$ band has $\sigma=N-2$.
(Note that, alternatively, for a different ratio $\kappa_0/\kappa_2$,
the $K=0_2$ band also can have $\sigma=N$ character as in~\cite{isa99}.)
In this case the diagonal $\kappa_0$ term in Eq.~(\ref{h1}) simply
shifts each band as a whole in accord with its $\sigma$ assignment.
All eigenstates of both $H_1$ and $H_2$
are mixed with respect to $O(5)$.
This is demonstrated in Fig.~3 for the $L=0,2$ members
of the respective ground bands.
The observed $\Delta\tau=\pm1,\pm3$ mixing
is generated by the $\kappa_2$ term in $H_1$ (\ref{h1}),
and by the $\Pi^{(2)}\cdot(d^{\dagger}\tilde{d})^{(2)}$ term
in $H_2$ (\ref{mult}),
which are both $(3,0)$ tensors with respect to $O(5)$.
The combined results of Figs.~2 and~3 constitute a direct proof
that $H_2$ possesses a generalized $O(6)$ PDS
which is distinct from the PDS of $H_1$.

To gain more insight into the underlying band structure of $H_2$
we perform a band-mixing calculation
by taking its matrix elements between large-$N$ intrinsic states.
The latter are obtained in the usual way
by replacing a condensate boson in $\vert c;\,N \rangle$ (\ref{cond})
with orthogonal bosons
$b^\dagger_\beta=(d^\dagger_0-s^\dagger)/\sqrt{2}$
and $d^\dagger_{\pm2}$
representing $\beta$ and $\gamma$ excitations, respectively.
By construction, the intrinsic state for the ground band of $H_2$,
$\vert K=0_1 \rangle = \vert c;\, N\rangle$, is decoupled.
For the lowest excited bands we find
\ba
\vert K=0_2\, \rangle &=&
A_{\beta}\,\vert\beta\rangle +
A_{\gamma^2}\,\vert\gamma^2_{K=0}\,\rangle +
A_{\beta^2}\,\vert\beta^2\,\rangle ~,
\nonumber\\
\vert K=2_1\, \rangle &=&
A_{\gamma}\,\vert\gamma\,\rangle +
A_{\beta\gamma}\,\vert\beta\gamma\,\rangle ~.
\label{intstat}
\ea
Using the parameters of $H_2$ relevant to $^{162}$Dy (see Fig.~1),
we obtain that the $K=0_2$ band is composed of
$36.29\%$ $\beta$, $63.68\%$ $\gamma^2_{K=0}$,
and $0.03\%$ $\beta^2$ modes,
{\it i.e.}, it is dominantly a double-gamma phonon excitation
with significant single-$\beta$ phonon admixture.
The $K=2_1$ band is composed of
$99.85\%$ $\gamma$ and $0.15\%$ $\beta\gamma$ modes,
{\it i.e.}, it is an almost pure single-gamma phonon band.
An $O(6)$ decomposition of the intrinsic states in Eq.~(\ref{intstat})
shows that the $K=0_2$ intrinsic state
has components with $\sigma=N\,(86.72\%)$,
$\sigma=N-2\,(13.26\%)$, and $\sigma=N-4\,(0.02\%)$.
The $K=2_1$ intrinsic state
has $\sigma=N$ $(99.88\%)$ and $\sigma=N-2$ $(0.12\%)$.
These estimates are in good agreement
with the exact results mentioned above in relation to Fig.~2.

In Table~I we compare the presently known experimental $B(E2)$ values
for transitions in $^{162}$Dy
with the values predicted by $H_1$ and $H_2$
using the $E2$ operator
\ba
T^{(2)} &=&
e_{B}\left[\, 
\Pi^{(2)} + \chi\, (d^{\dagger}\tilde{d}\,)^{(2)}\,\right ] ~.
\label{e2}
\ea
Absolute $B(E2)$ values are known
for transitions within the $K=0_1$ ground band~\cite{helmer99}.
The experimental values for the $K=2_1\rightarrow K=0_1$ transitions
are deduced from measured branching ratios
together with the assumption of equal intrinsic quadrupole moments
of the two bands~\cite{warner88,warner02}.
The latter assumption is satisfied
by the calculated $E2$ rates to within about 10\%.
The parameters $e_{B}$ and $\chi$ in Eq.~(\ref{e2})
are fixed for each Hamiltonian
from the empirical $2^+_{K=0_1}\rightarrow 0^+_{K=0_1}$
and $2^+_{K=2_1}\rightarrow 0^+_{K=0_1}$ $E2$ rates.
The $B(E2)$ values predicted by $H_1$ and $H_2$
for $K=0_1\rightarrow K=0_1$ and $K=2_1\rightarrow K=0_1$ transitions
are very similar and agree well with the measured values.
On the other hand, their predictions for interband transitions
from the $K=0_2$ band are very different.
For $H_1$, the $K=0_2\rightarrow K=0_1$ and $K=0_2\rightarrow K=2_1$
transitions are comparable and weaker than $K=2_1\rightarrow K=0_1$.
This can be understood
if we recall the $O(6)$ assignments for the bands of $H_1$
[$K=0_1,\,2_1$: $\sigma=N$; $K=0_2$: $\sigma=N-2$]
and the $E2$ selection rules of $\Pi^{(2)}\,(\Delta\sigma=0)$
and $(d^{\dagger}\tilde{d}\,)^{(2)}\,(\Delta\sigma=0,\pm 2)$,
which imply that in this case
only the $(d^{\dagger}\tilde{d}\,)^{(2)}$ term
contributes to interband transitions from the $K=0_2$ band.
In contrast, for $H_2$,
$K=0_2\rightarrow K=2_1$ and $K=2_1\rightarrow K=0_1$ transitions
are comparable and stronger than $K=0_2\rightarrow K=0_1$.
This behavior is a consequence of the underlying band structure
discussed above,
and the fact that
$\langle K=0_2\, \vert \Pi^{(2)}_{0} \vert\, K=0_1\rangle =0$,
while both terms in Eq.~(\ref{e2})
contribute to $\Delta K=2$ interband $E2$ intrinsic matrix elements.
Recently, the $B(E2)$ ratios
$R_1={B(E2;\,0^+_{K=0_2}\rightarrow 2^+_{K=2_1})\over
B(E2;\,0^+_{K=0_2}\rightarrow 2^+_{K=0_1})} = 10(5)$
and $R_2={B(E2;\,2^+_{K=0_2}\rightarrow 4^+_{K=0_1})\over
B(E2;\,2^+_{K=0_2}\rightarrow 0^+_{K=0_1})} = 65(28)$ 
have been measured \cite{zamfir99}.
The corresponding predictions are $R_1= 0.90$, $R_2=3.76$ for $H_1$
and $R_1=75.09$, $R_2=3.77$ for $H_2$,
and are at variance with the observations. 
However, as noted in \cite{zamfir99}, the empirical value of $R_2$ deviates 
`beyond reasonable expectations' from the Alaga rules value $R_2=2.57$. 
A measurement of absolute $B(E2)$ values for these transitions
is highly desirable to clarify the origin of these discrepancies.

To summarize, we have introduced
the concept of a generalized partial dynamical symmetry.
An illustration was given for the interacting boson model
by introducing Hamiltonians
that are not invariant under $O(6)$
but have a subset of solvable eigenstates with good $O(6)$ symmetry,
while other states are mixed.
None of the states conserves the $O(5)$ symmetry.
This novel intermediate-symmetry structure
has features relevant to axially deformed nuclei
whose $\Delta K=2$ interband transitions from the $K=2_1,\,0_2$ bands
are stronger than $\Delta K=0$ interband transitions
from the $K=0_2$ band to the ground band.

This work was supported in part (A.L.) by the Israel Science Foundation.

\renewcommand{\baselinestretch}{1.5}
\begin{table}
\caption[]{Calculated and observed~\protect\cite{helmer99,warner88}
$B(E2)$ values (in $e^2b^2$) for $^{162}$Dy.
The $E2$ parameters in Eq.~(\ref{e2}) are
$e_{B}=0.138$ $(0.127)$ $eb$ and $\chi=-0.235$ $(-0.557)$
for $H_1$ $(H_2)$.
\normalsize}
\vskip 10pt
\begin{tabular}{llll|llll}
Transition & $H_{1}$ & $H_{2}$ & Expt. & 
Transition & $H_{1}$ & $H_{2}$ & Expt. \\
\hline
$2^{+}_{K=0_1}\rightarrow 0^{+}_{K=0_1}$  & 1.07   & 1.07   & 1.07(2) &
$2^{+}_{K=2_1}\rightarrow 0^{+}_{K=0_1}$  & 0.024  & 0.024  & 0.024(1)  \\
$4^{+}_{K=0_1}\rightarrow 2^{+}_{K=0_1}$  & 1.51   & 1.52   & 1.51(6) &
$2^{+}_{K=2_1}\rightarrow 2^{+}_{K=0_1}$  & 0.038  & 0.040  & 0.042(2)  \\
$6^{+}_{K=0_1}\rightarrow 4^{+}_{K=0_1}$  & 1.63   & 1.65   & 1.57(9) &
$2^{+}_{K=2_1}\rightarrow 4^{+}_{K=0_1}$  & 0.0024 & 0.0026 & 0.0030(2) \\
$8^{+}_{K=0_1}\rightarrow 6^{+}_{K=0_1}$  & 1.66   & 1.68   & 1.82(9)   &
$3^{+}_{K=2_1}\rightarrow 2^{+}_{K=0_1}$  & 0.042  & 0.043  &           \\
$10^{+}_{K=0_1}\rightarrow 8^{+}_{K=0_1}$ & 1.64   & 1.67   & 1.83(12)  &
$3^{+}_{K=2_1}\rightarrow 4^{+}_{K=0_1}$  & 0.022  & 0.023  &           \\
$12^{+}_{K=0_1}\rightarrow 10^{+}_{K=0_1}$& 1.59   & 1.63   & 1.68(21)  &
$4^{+}_{K=2_1}\rightarrow 2^{+}_{K=0_1}$ & 0.0121  & 0.0114 & 0.0091(5)  \\
 & & & &
$4^{+}_{K=2_1}\rightarrow 4^{+}_{K=0_1}$ & 0.045  & 0.047  & 0.044(3)   \\
$0^{+}_{K=0_2}\rightarrow 2^{+}_{K=0_1}$ & 0.0016 & 0.0023 &         & 
$4^{+}_{K=2_1}\rightarrow 6^{+}_{K=0_1}$ & 0.0059 & 0.0061 & 0.0063(4)  \\
$0^{+}_{K=0_2}\rightarrow 2^{+}_{K=2_1}$ & 0.0014 & 0.1723 &         &
$5^{+}_{K=2_1}\rightarrow 4^{+}_{K=0_1}$ & 0.034  & 0.033  & 0.033(2)   \\
$2^{+}_{K=0_2}\rightarrow 0^{+}_{K=0_1}$ & 0.0002 & 0.0004 &         &
$5^{+}_{K=2_1}\rightarrow 6^{+}_{K=0_1}$ & 0.029  & 0.031  & 0.040(2)   \\
$2^{+}_{K=0_2}\rightarrow 2^{+}_{K=0_1}$ & 0.0004 & 0.0005 &         &
$6^{+}_{K=2_1}\rightarrow 4^{+}_{K=0_1}$ & 0.0084 & 0.0072 & 0.0063(4)  \\
$2^{+}_{K=0_2}\rightarrow 2^{+}_{K=2_1}$ & 0.0003 & 0.0369 &         &
$6^{+}_{K=2_1}\rightarrow 6^{+}_{K=0_1}$ & 0.045  & 0.047  & 0.050(4)   \\
\end{tabular}
\end{table}

\clearpage

\renewcommand{\baselinestretch}{1.7}
\begin{figure}
\caption{Experimental spectra (EXP)
of $^{162}$Dy~\protect\cite{helmer99,zamfir99}
compared with calculated spectra of
$H_1+\lambda_1 L\cdot L$, Eq.~(\ref{h1}), and
$H_2+\lambda_2 L\cdot L$, Eq.~(\ref{h2}), with parameters (in keV)
$\kappa_0=8$, $\kappa_2=1.364$, $\lambda_1=8$ and
$h_0=28.5$, $h_2=6.3$, $\lambda_2=13.45$
and boson number $N=15$.}
\end{figure}

\begin{figure}
\caption{$O(6)$ decomposition of wave functions of states in the bands 
$K=0_1,\,2_1,\,0_2,\,(L=K^{+})$, and $K=2_3,\,(L=3^{+})$, for 
$H_1$ (upper portion) and $H_2$ (lower portion).}
\end{figure}

\begin{figure}
\caption{$O(5)$ decomposition of wave functions of the $L=0,\,2$ states
in the ground band ($K=0_1$) of $H_1$ (upper portion) and
$H_2$ (lower portion). All states have $\sigma=N$.}
\end{figure}

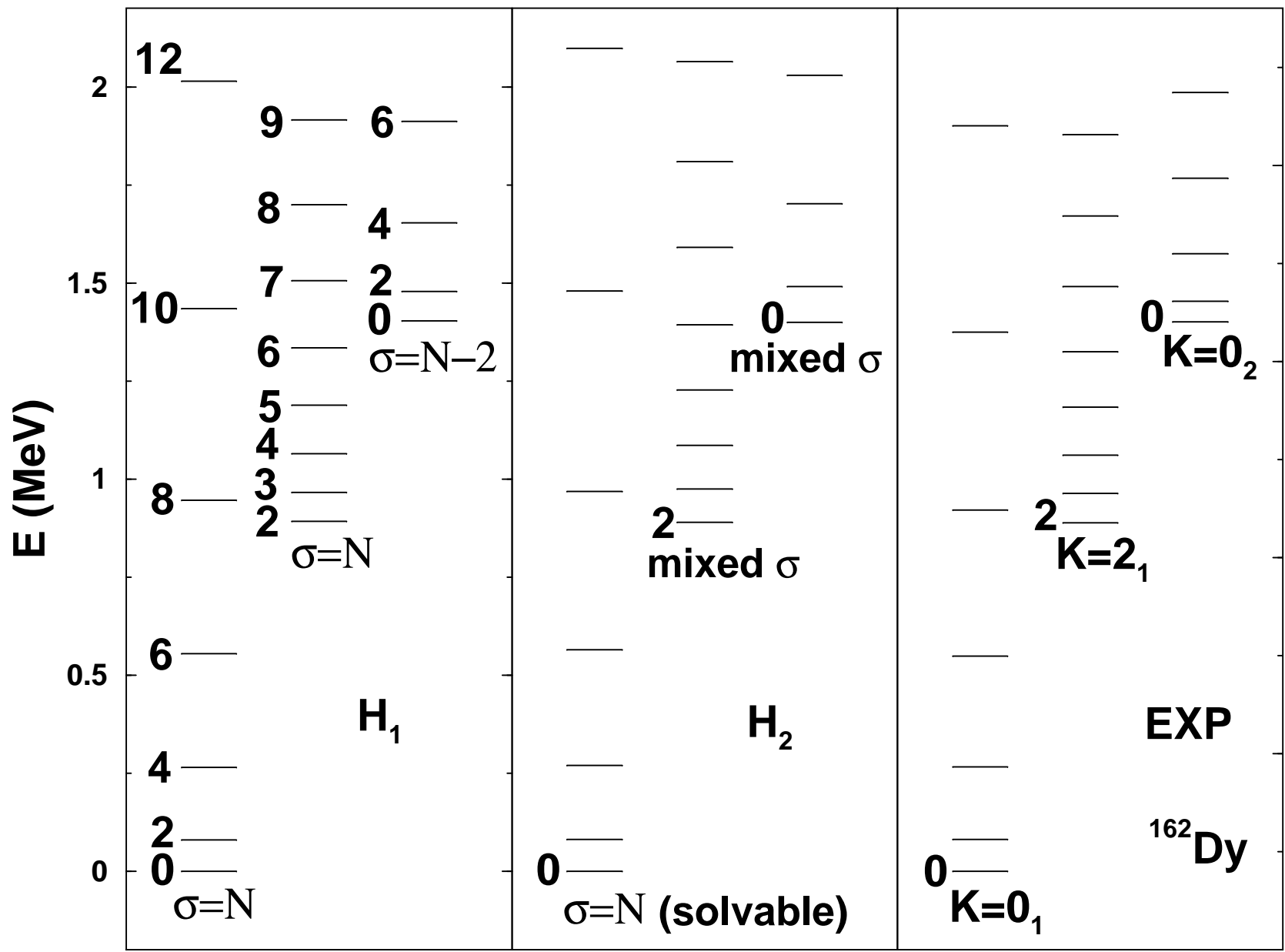
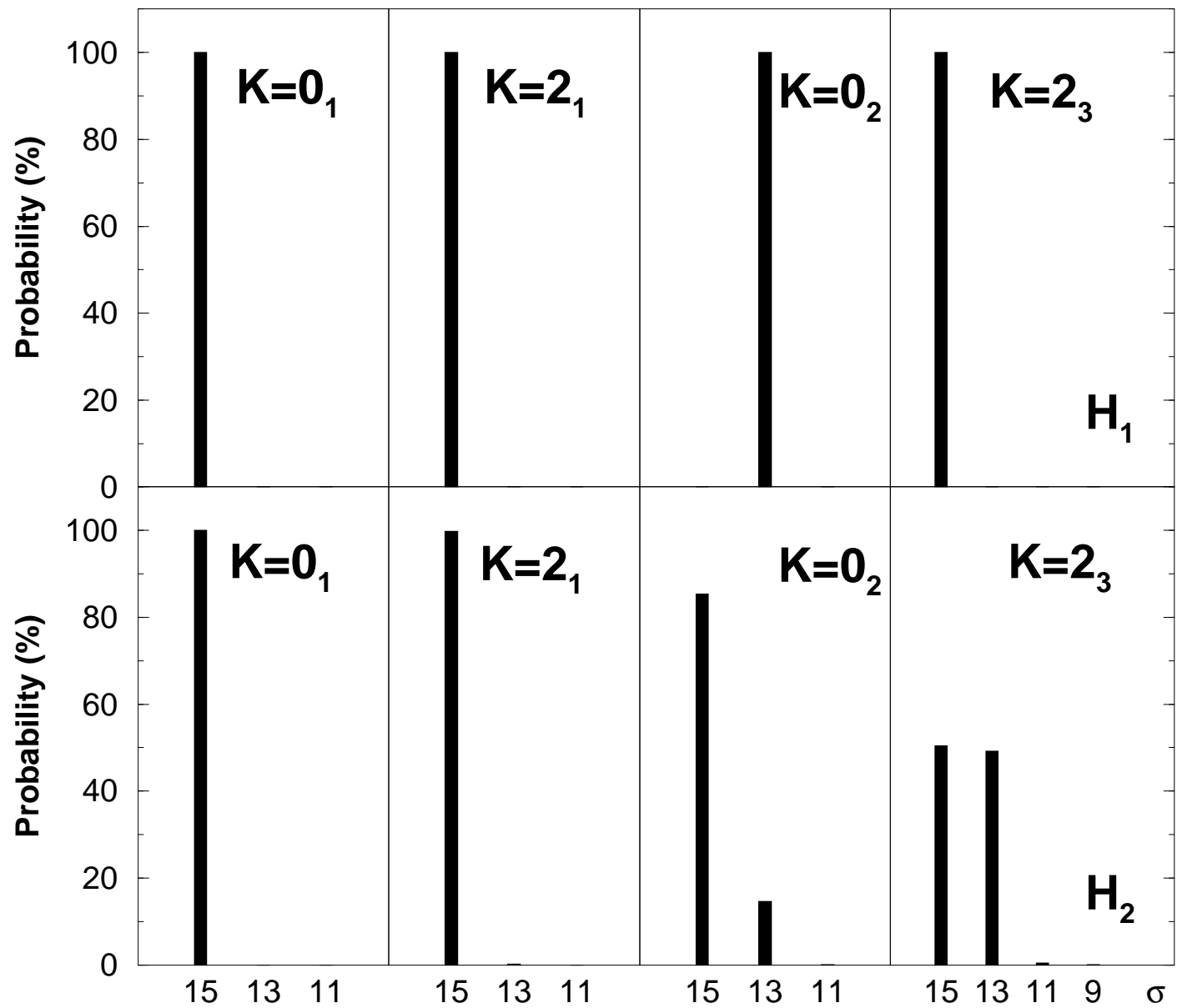
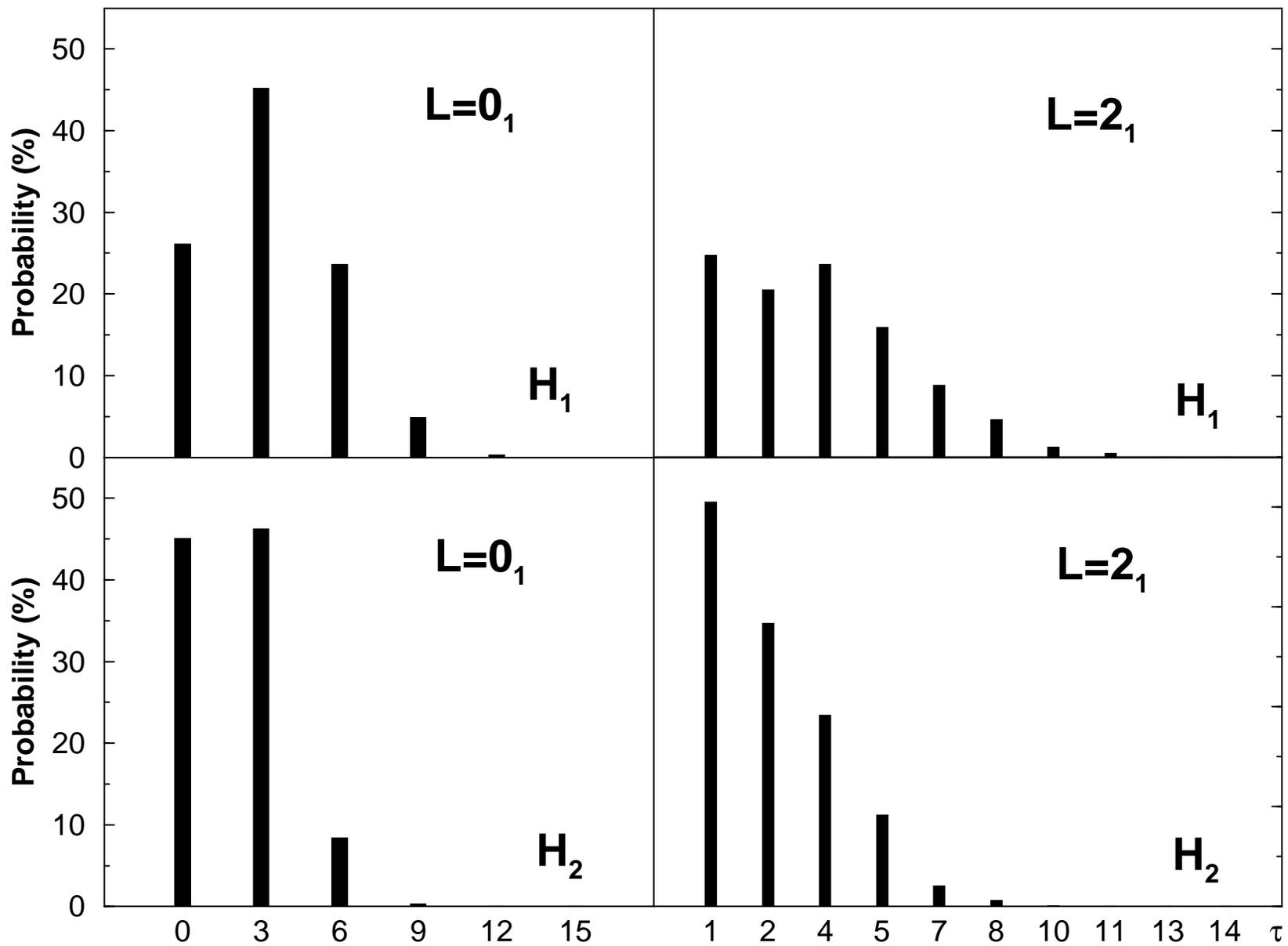
\end{document}